\documentclass[%
 reprint,
 groupedaddress,
 nofootinbib,
 nobibnotes,
 amsmath,amssymb,
 aps,
 prd,
]{revtex4-1}
\usepackage{graphicx}
\usepackage{dcolumn}
\usepackage{bm}
\usepackage{booktabs}
\usepackage{amsmath,amssymb,amsbsy,amstext, amsthm, amsfonts}
\usepackage{color}
\usepackage{slashed}
\usepackage[dvipsnames]{xcolor}
\usepackage{tikz}
\usepackage{dsfont}
\usepackage{verbatim}
\usepackage{comment}
\usepackage[utf8]{inputenc} 
\usepackage{ragged2e}
\usepackage{physics}
\usepackage[hidelinks]{hyperref}

\begin{document}
\preprint{APS/123-QED}

\title{Primordial Gravitational Waves From Black Hole Evaporation\\ in Standard and Non-Standard Cosmologies}

\author{Aurora Ireland$^{1}$}
\thanks{ORCID: https://orcid.org/0000-0001-5393-0971}

\author{Stefano Profumo$^{2,3}$}
\thanks{ORCID: http://orcid.org/0000-0002-9159-7556}

\author{Jordan Scharnhorst$^{2,3}$}
\thanks{ORCID: http://orcid.org/0000-0001-7112-3313}

\affiliation{$^1$University of Chicago, Department of Physics, Chicago, IL 60637, USA}
\affiliation{$^2$Department of Physics, University of California, Santa Cruz, CA 95060, USA}
\affiliation{$^3$Santa Cruz Institute for Particle Physics, University of California, Santa Cruz, CA 95060, USA}

\date{\today}

\begin{abstract}
Gravitons radiated from light, evaporating black holes contribute to the stochastic background of gravitational waves. The spectrum of such emission depends on both the mass and the spin of the black holes, as well as on the redshifting that occurs between the black hole formation and today. This, in turn, depends on the expansion history of the universe, which is largely unknown and unconstrained at times prior to the synthesis of light elements. Here, we explore the features of the stochastic background of gravitational waves from black hole evaporation under a broad range of possible early cosmological histories. The resulting gravitational wave signals typically peak at very high frequencies, and offer opportunities for proposed ultra-high frequency gravitational wave detectors. Lower-frequency peaks are also possible, albeit with a suppressed intensity that is likely well below the threshold of detectability. We find that the largest intensity peaks correspond to cosmologies dominated by fluids with equations of state that have large pressure-to-density ratios. Such scenarios can be constrained on the basis of violation of $\Delta N_{\rm eff}$ bounds.
\end{abstract}


\maketitle


\section{Introduction}
Black holes emit quasi-thermal radiation via the well-known process of Hawking evaporation \cite{Hawking:1974rv,Hawking:1975vcx}, through which they evaporate at calculable rates into all physical degrees of freedom with mass around or below the associated black hole temperature \cite{Hawking:1975vcx}. These degrees of freedom include gravitons; as such, black hole evaporation directly produces gravitational radiation, as pointed out long ago \cite{Page:1976ki}. This possibility is especially intriguing for light black holes of non-stellar, primordial origin --- primordial black holes, or PBH (see \cite{Carr:2021, Escriva:2022duf} for  recent reviews). In fact, this gravitational wave signal provides one of the few ways, if not {\em the only way} to probe light PBH with masses $M \lesssim 5 \times 10^8 \, \text{g}$, which evaporate prior to big bang nucleosynthesis (BBN) and are otherwise completely unconstrained.

Ref.~\cite{Dolgov:2000} and \cite{Anantua:2008} first studied the production of a stochastic gravitational wave background from light, evaporating PBH, including the possibility of an early matter domination (EMD) phase. We note, however, that both these studies neglected the black hole angular momentum and its evolution, as well as the corresponding large deviations from blackbody emission \cite{Page:1976ki}.  Ref.~\cite{Dolgov:2011} studied gravitational wave production from a number of mechanisms, including mergers and hawking evaporation, but computed the stochastic background of gravitational waves from evaporation while assuming both instantaneous black hole decay and a blackbody spectrum -- however, as we show below, these assumptions are inadequate. Finally, the calculations in Ref.~\cite{Dong:2015} included the effects of angular momentum and studied gravitational wave production from close-to-extremal Kerr black holes in a standard cosmological setting.

As we discuss in detail below, in light of the recent developments in the literature, the generic expectation for the gravitational wave background produced by evaporating {\em Schwarzschild} black holes with a {\em standard cosmological history} is two-fold: 
\begin{enumerate}
\item The peak frequency for gravitational wave emission is (see Eq.~(\ref{eq:fpeak}) below) $f_{\rm peak} \simeq (1.8 \times 10^{16} \, \text{Hz}) ( M/10^5 \, \text{g})^{1/2}$, and thus, even for very light black holes with masses close to the Planck scale, the signal is at ultra-high frequencies. 

\item The peak gravitational wave (GW) emission has an absolute maximum energy density $\Omega_{\rm GW} h^2 \big|_{\rm peak} \simeq 4.2 \times 10^{-7}$ (see Eq.~(\ref{eq:omegagwmax}) below). As such, the gravitational wave emission is possibly large enough to be constrained by measurements of the number of relativistic degrees of freedom (see e.g. the recent detailed study presented in Ref.~\cite{Arbey:2021}), but likely not detectable neither by currently operating gravitational wave telescopes nor by planned high-frequency detectors (see \cite{Aggarwal:2020} for a review).
\end{enumerate}

Here, we examine how assumptions about the very early universe affect the expectations summarized above. Firstly, the peak gravitational wave frequency depends critically on how the emitted gravitons redshift, especially at very early times. Secondly, both the peak amplitude and location depend quite sensitively on the spin of the radiating black hole population (see e.g. \cite{Arbey:2021}). In particular, as pointed out above, spinning black holes radiate gravitons both more abundantly and peaking at lower frequencies \cite{Page:1976a}. Finally, the cosmological history drastically affect the maximal gravitational wave intensity, to the level of enhancing it by several orders of magnitude, making graviton emission a prime target for future high-frequency gravitational wave searches.

Using state-of-the-art tools such as the {\tt BlackHawk} package \cite{Arbey:2021}, we explore the features of the stochastic background of gravitational waves stemming from Hawking evaporation of light PBH with non-standard, non-radiation-dominated cosmologies at early times. An especially well-motivated scenario is the possibility of an early phase of matter domination \cite{Dolgov:2000,Arbey:2021}; more generically, prior to BBN, the universe's energy density could have been dominated by a species $\phi$ with a generic equation of state $P_\phi=w_\phi\rho_\phi$.

The remainder of this study is as follows: in Sec.~\ref{sec:pbhevolution} we review the formation and evolution of PBH, especially for the case of near-extremal spin. Sec.~\ref{sec:gw} examines, both numerically and analytically, the gravitational wave production from Hawking evaporation of gravitons, and elucidates the impact of the blackbody and instant decay approximations. Sec.~\ref{sec:nonstandard} introduces non-standard cosmological histories, and examines the impact thereof on gravitational waves from Hawking evporation of gravitons. The final Sec.~\ref{sec:conclusions} discusses observational prospects and constraints, and concludes. 

\section{PBH Formation \& Evaporation}\label{sec:pbhevolution}

\subsection{PBH Formation}

PBHs can form from a variety of mechanisms in the early universe, though most involve the collapse of matter overdensities seeded by inflation or topological defects \cite{Escriva:2022duf}. The resulting mass spectrum and energy density then depend on the time of formation, reheating temperature, and shape of the inflationary potential. In this work, we consider, for simplicity, monochromatic mass spectra without loss of generality, and leave the initial mass fraction at formation $\Omega_{\rm BH}^i \equiv \rho_{\rm BH}/\rho_{\rm tot}$ as a free parameter, subject to constraints.

The mass $M$ of a PBH formed at time $t_i$ roughly corresponds to the mass contained within a Hubble horizon at $t_i$, $M_{\rm hor} = \frac{4 \pi}{3} H^{-3} \rho$. Using the fact that $\rho = \frac{3 M_{\rm Pl}^2}{8 \pi} H^2$ and $H \sim \frac{1}{2 t}$ during radiation domination, one can then relate the initial PBH mass to cosmological time in radiation-domination as:

\begin{equation}
    M \sim 10^{10}\ \left( \frac{t_i}{10^{-28}\, \text{s}} \right)\, \text{g}.
\end{equation}
The abundance of PBHs with initial mass greater than $\sim 5 \times 10^8 \, \text{g}$ is strongly constrained by evaporation, gravitational lensing, gravitational waves from binary mergers, dynamical effects, accretion, and large scale structure (see e.g. Fig. 3 of \cite{Green:2020} for a summary). However, the abundance of PBHs which decay before BBN, corresponding roughly to initial masses $\lesssim 5 \times 10^8 \, \text{g}$, are largely unconstrained, absent specific assumptions \cite{Carr:2021}.

Note that a Schwarzschild black hole of mass $M$ has a lifetime \cite{Carr:2021}:
\begin{equation}
    \tau_{\rm BH}\simeq 407 \left(\frac{M}{10^{10}\ \rm g}\right)^3\ {\rm s},
\end{equation}
so PBH evaporating before BBN must satisfy:
\begin{equation}
\frac{t_i}{1\ \rm s}+4\times 10^{86}\left(\frac{t_i}{10^{-28}\ {\rm s}}\right)^3\lesssim 1 .
\end{equation}
Here, the second term always dominates the inequality for  $t>t_{\rm Pl}$, with $t_{\rm Pl}=\sqrt{G\hbar/c^5}$ the Planck time, so here we assume that the PBH mass $M<5\times 10^8$ g, for which the lifetime is shorter than  1 s.  
Note that the lifetime of Kerr black holes is reduced by around a factor of one half for near-maximal spinning black holes, leaving these estimates unchanged.

Kerr black holes, which have non-vanishing angular momentum $J$ and preferentially emit higher spin particles, like the spin-2 graviton, are especially interesting in the present analysis. In Boyer-Lindquist coordinates, the geometry of a Kerr black hole is described by the metric \cite{Kerr:1963}
\begin{equation}\label{KNmetric}
\begin{split}
	ds^2 = & - \frac{\Delta}{\rho^2} (dt - \alpha \sin^2 \theta \, d\phi)^2 + \frac{\rho^2}{\Delta} dr^2 \\
	& + \rho^2 d\theta^2 + \frac{\sin^2 \theta}{\rho^2} \left[ (r^2+\alpha^2) d\phi - \alpha dt \right]^2 \,,
\end{split}
\end{equation}
where $M_{\rm Pl} = \sqrt{G \hbar/c^3}=1.22 \times 10^{19}$ GeV is the Planck mass\footnote{Hereafter we shall use natural units $\hbar=c=1$.}, $M$ is the black hole mass, $\alpha = J/M$ is the spin parameter, and we have defined:
\begin{equation}
    \rho^2 = r^2 + \alpha^2 \cos^2 \theta \,, \,\,\,\,\,\, \Delta = r^2 + \alpha^2 - \frac{2 M r}{M_{\rm Pl}^2}  \,.
\end{equation}
It will also be convenient to define the dimensionless spin parameter $\alpha_\star = \frac{M_{\rm Pl}^2}{M^2} J$, which can take values $\alpha_\star \in [0,1]$, with $\alpha_\star = 1$ corresponding to the extremal case. Kerr black holes have two horizons $r_\pm$, with the outer horizon located at:
\begin{equation}
    r_+ = \frac{M}{M_{\rm Pl}^2}(1 + \sqrt{1 - \alpha_\star^2}) \,.
\end{equation}
The Hawking temperature associated with this horizon is:
\begin{equation}\label{temperature}
    T_{\rm BH} = \frac{M_{\rm Pl}^2}{4 \pi M} \frac{\sqrt{1-\alpha_\star^2}}{1 + \sqrt{1 - \alpha_\star^2}} \,.
\end{equation}
Note that this reduces to the temperature of the Schwarzschild black hole $T = \frac{M_{\rm Pl}^2}{8 \pi M}$ in the limit of vanishing spin $\alpha_\star \rightarrow 0$, and tends to 0 in the extremal limit $\alpha_\star \rightarrow 1$. The Kerr black hole is also characterized by an angular velocity $\Omega_{\rm BH}$ given by:
\begin{equation}\label{angmom}
    \Omega_{\rm BH} = \frac{M_{\rm Pl}^2}{2 M} \frac{\alpha_\star}{1 + \sqrt{1- \alpha_\star^2}} \,.
\end{equation}

\subsection{Hawking Evaporation \& Black Hole Evolution }

The flux spectrum for the emission of a single particle degree of freedom\footnote{In order to obtain the total flux per particle species $i$, one would sum over the polarization and charge degrees of freedom.} of a species $i$ with frequency $\omega$ and spin $s$ is \cite{Hawking:1975,Page:1976a,Page:1976b}
\begin{equation}
    \frac{dN_i(\omega)}{dt} = \sum_{\ell, m} \frac{\sigma_{\ell m}^{(s)}(\omega)}{e^{(\omega - m \Omega)/T_{\rm BH}} - (-1)^{2s}} \frac{d^3k}{(2\pi)^3} \,,
\end{equation}
where the sum runs over the total $\ell$ and axial $m$ angular momenta of the emitted mode\footnote{We neglect here the effect of the charge of the particle on $\sigma_{\ell m}^{(s)}$}. This spectrum is almost that of a perfect blackbody, with the deviation captured by the ``greybody factor" $\sigma_{\ell m}^{(s)}(\omega)$, which is related to the probability that a given mode will be able to surmount the gravitational potential barrier and escape to spatial infinity. 

Since the emission of particles with masses greater than the black hole temperature is exponentially suppressed, it often suffices to include in the sum for the total flux only those degrees of freedom lighter than the black hole. In this case we can evaluate the phase space factor to obtain the simplified expression for the emission of a massless degree of freedom per frequency interval
\begin{equation}\label{flux}
    \frac{dN_i}{dt d\omega} = \frac{1}{2\pi} \sum_{\ell, m} \frac{\Gamma_{\ell m}^{(s)}(\omega)}{e^{(\omega - m \Omega)/T_{\rm BH}} - (-1)^{2s}} \,,
\end{equation}
where $\Gamma_{\ell m}^{(s)} = \frac{\omega^2}{\pi} \sigma_{\ell m}^{(s)}$ is the absorption probability\footnote{The absorption probability $\Gamma_{\ell m}^{(s)}(\omega) = \frac{\omega^2}{\pi} \sigma_{\ell m}^{(s)}$ is sometimes referred to in the literature as the ``greybody factor". We refrain from doing so here to avoid confusion.}. 

\begin{figure}[t]
\centering
\includegraphics[width=0.45\textwidth]{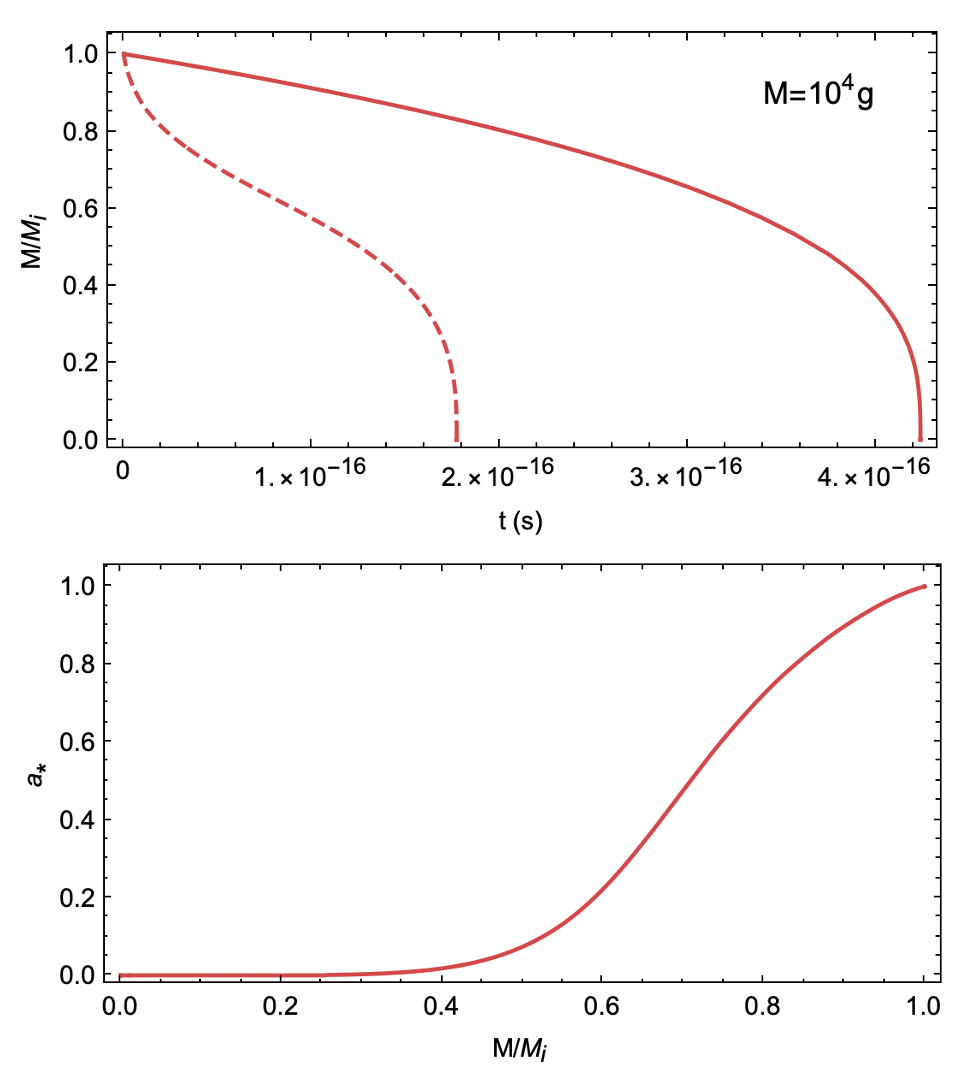}
\hspace{2mm}
\caption{Top: Evolution of black hole mass as a function of time for the case of a Schwarzschild black hole (solid red line) as compared with a near extremal ($\alpha_\star=0.999$) Kerr black hole (dashed line). The lifetime of a rapidly spinning black hole is reduced by a factor $\sim 2$. $M_i=10^4$ g is taken as a benchmark, but the above behavior is generic. Bottom: Evolution of dimensionless spin parameter $\alpha_\star$ as a function of $M/M_i$.}
\label{lifeevolutions}
\end{figure}

Each emitted particle carries off units of energy $\omega$ and of angular momentum $m$ about the black hole axis. Note that $m \Omega$ acts as an effective chemical potential, biasing the emission of particles whose angular momentum is aligned with that of the black hole. In this manner, a black hole sheds both mass and angular momentum, and evolves toward a non-rotating state. The power emitted in a given frequency interval per particle degree of freedom is:
\begin{equation}\label{Eflux}
    \frac{dE_i}{dt d\omega} = \frac{1}{2\pi} \sum_{\ell, m} \frac{\omega \Gamma_{\ell m}^{(s)}(\omega)}{e^{(\omega - m \Omega)/T_{\rm BH}} - (-1)^{2s}} \,.
\end{equation}
This enters into the rate at which the black hole loses mass as \cite{Page:1976a}
\begin{equation}\label{massrate}
    \frac{dM}{dt} = - \sum_i \int d\omega \left( \frac{d E_i}{dt d\omega} \right) \,,
\end{equation}
where the sum runs over all degrees of freedom $i$ emitted by the black hole. Similarly, angular momentum is lost at a rate:
\begin{equation}\label{Jrate}
    \frac{dJ}{dt} = - \sum_i \int d\omega \left( m \frac{d N_i}{dt d\omega} \right) \,.
\end{equation}
In practice, it is convenient to track the evolution by introducing the dimensionless ``Page factors'' $f$ and $g$, defined implicitly via \cite{Page:1976b}:
\begin{subequations}\label{pagefactors}
\begin{equation}
    f(M,\alpha_\star) = - M^2 \frac{dM}{dt} \,,
\end{equation}
\begin{equation}
    g(M,\alpha_\star) = - \frac{M}{\alpha_\star} \frac{dJ}{dt} \,.
\end{equation}
\end{subequations}
Given explicit forms for the greybody factors of all relevant particle species, the contributions to $f$ and $g$ from each species can be numerically evaluated and the values tabulated. These can then be interpolated for the functions $f(M,\alpha_\star)$ and $g(M,\alpha_\star)$, from which one can solve Eq. (\ref{pagefactors}) to obtain the black hole mass and angular momentum as a function of time. Sample evolutions are shown Fig. \ref{lifeevolutions}. 

Fig. \ref{lifeevolutions} illustrates that the black hole mass remains roughly constant near its initial value until the very end of its lifetime, at which point it falls off dramatically. Angular momentum serves to reduce the black hole lifetime by an $\mathcal{O}(1)$ factor, with rapidly spinning black holes evaporating more quickly than their Schwarzschild counterparts. Because angular momentum decreases more rapidly than mass does, Kerr black holes finish shedding angular momentum and transition to a Schwarzschild phase before evaporating completely, as can be seen in Fig. \ref{lifeevolutions}. Once the black hole has spun down (seen in the lower panel to occur around $M/M_i \sim 0.4$), it evolves identically to the Schwarzschild case, as can be seen by the identical slopes at low $M/M_i \lesssim 0.4$ in the upper panel.

\subsection{Greybody Factors}

While approximating the spectrum as a perfect blackbody usually suffices for estimating the black hole lifetime, computing the gravitational wave signal will require a precise knowledge of the greybody factors. In general, these depend on the frequency, angular momentum, and spin of the emitted particle species, as well as the structure of spacetime about the black hole. Computation of the greybody factors is quite non-trivial and involves solving the relevant equation of motion for a given particle species on a curved black hole background with appropriate boundary conditions --- the Teukolsky equations \cite{Teukolsky:1972,Teukolsky:1974}. The absorption probability for a given mode, and thereby the greybody factor, is then determined by taking ratios of the amplitudes for incoming and outgoing waves at infinity.

The Teukolsky equations generically need to be solved numerically\footnote{Numerical solutions typically have issues with convergence at the black hole horizon and spatial infinity. For this reason, {\tt BlackHawk} employs the methods outlined in Ref.~\cite{Chandrasekhar:1975,Chandrasekhar:1976a,Chandrasekhar:1976b,Chandrasekhar:1977} to transform the Teukolsky equations into Schr{\"o}dinger-like wave equations with appropriately chosen short-range potentials.}, though analytic approximations exist in the low frequency limit, $M \omega/M_{\rm Pl}^2 \ll 1$. In particular for the $s=2$ graviton, the  greybody factor, summed over angles, in the low frequency limit reads \cite{Starobinsky:1974,Page:1976a}
\begin{equation}
    \sum_{\ell,m} \sigma^{(2)}_{\ell m} \xrightarrow[]{\omega \sim 0} \frac{16 A}{225} \left( 5 \frac{M_{\rm Pl}^2}{M^2} + \frac{5}{2} \alpha_\star^2 + \alpha_\star^4 \right) \left(\frac{M \omega}{M_{\rm Pl}^2} \right)^4 \,,
\end{equation}
where $A = 4\pi r_+^2$ is the black hole area. Note that this is highly suppressed, scaling with the frequency as $\omega^4$. In contrast, $\sum \sigma_{\ell m}^{(0)} \sim \omega^0$ for scalars, $\sum \sigma_{\ell m}^{(1/2)} \sim \omega^0$ for fermions, and $\sum \sigma_{\ell m}^{(1)} \sim \omega^2$ for vector bosons. The suppression at low frequencies in the graviton case can be understood by recognizing that the dominant contribution to $\sigma$ comes from the mode of lowest $\ell$, and since $\ell \geq s$ this is $\ell=2$ for the graviton. Meanwhile in the high frequency limit $M \omega/M_{\rm Pl}^2 \gg 1$, the greybody factors for all particle species approach the geometric optics limit, which is essentially the emitting area of the black hole.

\section{Standard Cosmological Evolution}\label{sec:gw}

\subsection{Analytical Estimates}

We are interested here in the present-day energy density in the form of gravitational waves from  evaporating PBHs, as parameterized by the spectral density parameter $\Omega_{\rm GW}$, defined as
\begin{equation}\label{def}
    \Omega_{\rm GW} = \frac{1}{\rho_{\rm crit}} \frac{d \rho_{\rm GW}}{d \ln f} \,,
\end{equation}
with $\rho_{\rm crit}$ the critical energy density today. Before turning to numerics, we will demonstrate how this can be computed starting from the instantaneous spectrum of graviton emission, $\frac{d E_{\rm grav}}{dt d\omega}$. For the sake of having analytic expressions, we will restrict ourselves to Schwarzschild black holes ($\alpha_\star = 0$) in the blackbody approximation, for which the greybody factor is simply the frequency-independent area of the black hole $\sum \sigma_{\ell m}^{(s)} = 4 \pi r_s^2$. We will also presume instantaneous decay, taking the black hole mass and temperature to be constants up until the moment of evaporation at $\tau_{\rm BH}$. These approximations will be all relaxed in the subsequent sections where we present our numerical results. 

Our starting point is the instantaneous energy flux expression of Eq. (\ref{Eflux}). Taking $\sum \sigma_{\ell m} = 4\pi r_s^2 = 16 \pi \frac{M^2}{M_{\rm Pl}^4}$ and multiplying by $g_i = 2$ for the two graviton polarizations, this becomes:
\begin{equation}
    \frac{dE_{\rm grav}}{dt d\omega} \simeq \frac{16}{\pi} \frac{M^2}{M_{\rm Pl}^4} \frac{\omega^3}{e^{\omega/T_{\rm BH}} - 1} \,.
\end{equation}
To obtain the rate of graviton emission for an entire population of evaporating PBHs, we multiply by the number density $n_{\rm BH}(t)$:
\begin{equation}
    \frac{d\rho_{\rm GW}}{dt d\omega} \simeq n_{\rm BH}(t) \frac{dE_{\rm grav}}{dt d\omega} \,.
\end{equation}
This should then be integrated over the black hole lifetime in order to determine the total amount of energy density in the form of gravitational waves at the time of evaporation. Let $t_i$ be the time of black hole formation, when graviton emission commences, and let $t_* = t_i + \tau_{\rm BH} \simeq \tau_{\rm BH}$ be the time of black hole evaporation, which, for a Schwarzschild black hole, is approximately
\begin{equation}
    \tau_{\rm BH} \simeq \frac{10240 \pi}{g_{\star,H}} \frac{M^3}{M_{\rm Pl}^4} \,,
\end{equation}
where $g_{\star,H} \simeq 108$ the number of effective degrees of freedom, since we restrict ourselves to light black holes evaporating before BBN, $M \lesssim 5 \times 10^8 \, \text{g}$. Since evaporation is occurring in an expanding universe, the black hole number density and graviton energy density and frequency are not fixed quantities, but rather experience cosmological redshift. In particular, they evolve as $n_{\rm BH} \sim a^{-3}$, $\rho_{\rm GW} \sim a^{-4}$, and $\omega \sim a^{-1}$, respectively. For ease of integration, we can isolate the time dependence by relating the graviton frequency and energy density to their values at the time of evaporation, which we denote by a star:
\begin{equation}
    \rho_{\rm GW} = \rho_{\rm GW}^* \left( \frac{a_*}{a} \right)^4 \,, \,\,\, \omega = \omega_* \left( \frac{a_*}{a} \right) \,.
\end{equation}
For the number density, it is more convenient to relate to the initial value 
\begin{equation}
    n_{\rm BH} = n_{{\rm BH}, i} \left( \frac{a_i}{a} \right)^3 \,,  
\end{equation}
which, in turn, can be related to the initial black hole mass and mass fraction $\Omega_{{\rm BH},i} = \rho_{{\rm BH},i}/\rho_{{\rm crit},i}$, presuming formation via the collapse of density perturbations in the radiation dominated early universe:
\begin{equation}\label{nBHi}
    n_{{\rm BH},i} = \frac{3 M_{\rm Pl}^6}{32 \pi M^3} \Omega_{{\rm BH},i} \,.
\end{equation}
Finally converting from frequency interval to logarithmic frequency interval $\frac{d}{d \ln \omega} = \omega \frac{d}{d\omega}$, the energy density in the form of gravitational waves at the time of evaporation is:
\begin{equation}\label{rhostar}
    \frac{d \rho_{\rm GW}^*}{d \ln \omega} \simeq \frac{16 n_{{\rm BH},i} M^2 \omega_*^4}{\pi M_{\rm Pl}^4} \int_{t_i}^{t_*} dt \, \frac{(a_i/a)^{3}}{e^{\omega_* a_*/a T_{\rm BH}} - 1} \,.
\end{equation}

As for the time dependence of the scale factor\footnote{When we turn to the numerical calculation, we will actually solve the Friedmann equations for the precise background evolution.}, initially during radiation domination it scales as $a \sim t^{1/2}$. If the initial energy density in black holes is sufficiently large and the black holes are sufficiently long lived, then they will eventually come to dominate the energy density of the universe at a time:
\begin{equation}
    t_{\rm eq} \simeq \left( \frac{1 - \Omega_{{\rm BH},i}}{\Omega_{{\rm BH},i}} \right)^2 \frac{M}{M_{\rm Pl}^2} \,.
\end{equation}
The condition on the initial energy density and mass for this to occur is $t_* > t_{\rm eq}$, or:
\begin{equation}\label{dominationcondition}
    \left( \frac{M}{10^5\, \text{g}} \right)^2 \left( \frac{\Omega_{{\rm BH},i}}{10^{-11}} \right)^2 \geq 1 \,.
\end{equation}
When this is satisfied, the universe will undergo a brief period of early matter domination from $t_{\rm eq}$ until $t_*$, during which the scale factor scales as $a \sim t^{2/3}$. Thus the scale factor appearing in Eq. (\ref{rhostar}) is:
\begin{equation}
    a(t) = 
    \begin{cases}
    a_i \left( \frac{t}{t_i} \right)^{1/2} & t \lesssim t_{\rm eq}\\
    a_i \left( \frac{t_{\rm eq}}{t_i} \right)^{1/2} \left( \frac{t}{t_{\rm eq}} \right)^{2/3} & t_{\rm eq} \lesssim t \lesssim t_* \,.
    \end{cases}
\end{equation}
One can also express $a(t)$ in terms of $a_*$ by noting that:
\begin{equation}
    \frac{a_i}{a_*} = \left( \frac{t_i}{t_{\rm eq}} \right)^{1/2} \left( \frac{t_{\rm eq}}{t_*} \right)^{2/3} \,.
\end{equation}

\begin{figure}[t]
\centering
\includegraphics[width=0.44\textwidth]{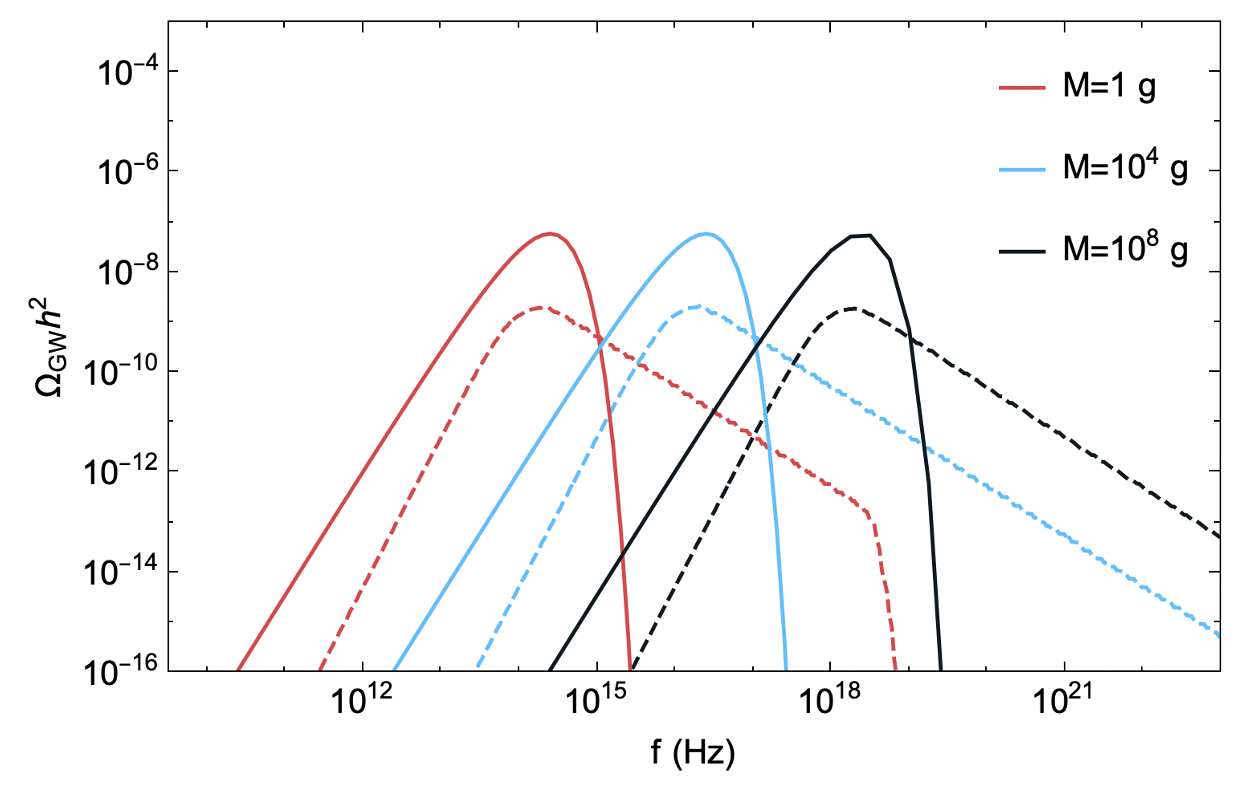}
\hspace{2mm}
\caption{Semi-analytic estimate (solid lines) for the spectral density parameter $\Omega_{\rm GW} h^2$ today, presuming a monochromatic spectrum of Schwarzschild black holes of initial mass $M$ and $\Omega_{{\rm BH},i}$ sufficiently large that the PBH eventually dominate the universe energy density, i.e. satisfying Eq. (\ref{dominationcondition}), such that the black holes come to dominate before decay. We work in the blackbody approximation and presume instantaneous decay. This can be compared with the exact numerical solution (dashed lines), for which these assumptions are relaxed.}
\label{analytic}
\end{figure}

Finally to translate the gravitational wave spectrum from evaporation to today, we need to account for the dilution of energy density and redshifting of frequency due to cosmological expansion. The energy density in the form of gravitational waves today is related to that at evaporation as
\begin{equation}\label{redshifttotoday}
    \frac{d \rho_{\rm GW}^0}{d \ln \omega_0} = \frac{d \rho_{\rm GW}^*}{d \ln \omega_*} \left( \frac{a_*}{a_0} \right)^4 \,,
\end{equation}
where $a_0 = a(t_0)$ is the scale factor today, which we take to be $a_0=1$. Explicit factors of the frequency appearing in this expression should be translated to their redshifted values today as $\omega_0 = \omega_* a_*$. Following black hole evaporation, the universe undergoes the usual epoch of radiation domination, and it is convenient to express the ratio of scale factors in terms of the plasma temperature and effective degrees of freedom in entropy, obtained via conservation of entropy $g_{\star,s} a^3 T^3 = \text{constant}$ as
\begin{equation}\label{astar}
    a_* = \left( \frac{g_{\star,s}(T_0)}{g_{\star,s}(T_{\rm RH})} \right)^{1/3} \frac{T_0}{T_{\rm RH}} \,,
\end{equation}
where $T_0 = 0.235 \, \text{meV}$ is the temperature of the cosmic microwave background (CMB) today and $g_{\star,s}(T_0)=3.91$. The reheating temperature for the SM plasma $T_{\rm RH} \equiv T(t_*)$ can be obtained by equating the energy density in the form of PBHs immediately before decay with the energy density in radiation immediately afterwards. Presuming black holes come to dominate prior to decay, this is approximately:
\begin{equation}
    T_{\rm RH} = 450 \left( \frac{g_\star(T_{\rm RH})}{106.75} \right)^{-1/4} \left( \frac{M}{10^5\, \text{g}} \right)^{-3/2} \text{GeV} \,.
\end{equation}
Substituting the spectral energy density today in the definition of Eq. (\ref{def}), we finally arrive at the prediction for the spectral density parameter today
\begin{equation}\label{OmegaGWanalytic}
    \Omega_{\rm GW} \simeq \frac{\Omega_{{\rm BH},i}}{H_0^2 M} \, \omega_0^4 \, I(\omega_0) \,,
\end{equation}
where $H_0 = 100 h \, \text{km} \cdot \text{s}^{-1} \cdot \text{Mpc}^{-1}$ is the Hubble rate, with $h\simeq 0.67-0.73$. The non-trivial frequency dependence lies in the integral
\begin{equation}
    I(\omega_0) = \int_{t_i}^{t_*} dt \, \frac{(a_i/a)^3}{e^{\omega_0/a T_{\rm BH}}-1} \,,
\end{equation}
which generically needs to be evaluated numerically for each $\omega_0$. Sample spectra are shown in Fig. \ref{analytic}.

The  gravitational wave spectrum from graviton production off of Hawking evaporation is almost thermal, but it features more power at low frequencies due to the redshifting of higher frequency modes into lower frequencies. The peak is generically located at very high frequencies --- far out of the range of current and near-future gravitational wave detectors. Decreasing the black hole mass shifts the peak frequency to lower values. This is because smaller black holes correspond to earlier formation times and evaporate more promptly, leading to a much longer period of cosmological redshift which serves to shift the spectrum to lower frequencies. Even saturating the mass bound by considering Planck scale black holes, however,  we remain outside of detector sensitivity.

By extremizing $\Omega_{\rm GW}$ with respect to $\omega_0$, one can show that the frequency today peaks at $\omega_{\rm peak} \simeq 2.8 a_* T_{\rm BH}$, or more explicitly
\begin{equation}\label{eq:fpeak}
    f_{\rm peak} \simeq (1.8 \times 10^{16} \, \text{Hz}) \left( \frac{M}{10^5 \, \text{g}} \right)^{1/2} \,,
\end{equation}
where we have converted to linear frequency $f = \omega/2\pi$. This is consistent with the peak positions in the sample spectra of Fig. \ref{analytic}. Evaluating $I(f_0)$ at the peak frequency, which dominates the contribution to the integral, one can show that the following empirical relation holds:
\begin{equation}
    I(f_{\rm peak}) \simeq (2.3 \times 10^{-33} \, \text{GeV}^{-1}) \left( \frac{M}{10^5 \, \text{g}} \right)^{-1} \Omega_{{\rm BH},i}^{-1} \,.
\end{equation}
Combining this with Eq. (\ref{OmegaGWanalytic}) evaluated at $f_{\rm peak}$, we find an estimate for the maximal value of the spectral density parameter which is, somewhat surprisingly, independent of both the initial black hole mass and fractional energy density, so long as these are sufficiently large that the black holes come to dominate the energy density of the universe before decay:
\begin{equation}\label{eq:omegagwmax}
    \Omega_{\rm GW} h^2 \big|_{\rm peak} \simeq 4.2 \times 10^{-7} \,.
\end{equation}
Note that we have extracted the reduced Hubble rate $h$ to alleviate the associated uncertainty in its value. This estimate is consistent with the peak amplitude shown in Fig. \ref{analytic} as well as with \cite{Dolgov:2011}. Comparing with current gravitational wave sensitivities (see e.g. Fig. A3 of \cite{Moore:2014}), we see that the magnitude of this signal at its peak is within reach of several current and proposed experiments; however this peak occurs as ultra-high frequencies far outside the current regime of observability. 

In the coming sections, we will see how a prolonged phase of early matter domination can actually give rise to extra cosmological redshift, which in turn serves to shift the peak emission to lower frequencies. This additional redshift, however, also has the effect of diluting the gravitational wave signal. In fact, since the energy density falls off as four powers of the scale factor $\rho_{\rm GW} \sim a^{-4}$ while the frequency scales with just one $f \sim a^{-1}$, the effect on $\Omega_{\rm GW}$ is much more significant. Thus,  to retain a detectable signal, one would need to enhance graviton emission. Recall that this estimate considered Schwarzschild black holes, for which only approximately 1\% of energy is emitted as gravitons. In contrast, Kerr black holes preferentially emit particles of higher spin, like the spin-2 graviton. For this reason, we consider Kerr black holes in the remainder of this work.

Finally by comparing with the exact spectra, obtained via {\tt BlackHawk} and shown also in Fig. \ref{analytic} as dashed lines, we see that the blackbody and instantaneous decay approximations are not adequate, although the peak positions coincide almost perfectly (which makes sense as the graviton peak frequency is largely fixed by the black hole temperature independently of greybody factors). As for the peak height, however, we see that the semi-analytic calculation overestimates the amplitude by several orders of magnitude due to the neglect of greybody factors, which otherwise suppress higher angular moments. Interestingly, however, the peak in the case of near-extremal Kerr black holes nearly matches the blackbody Schwarzschild estimate. Finally, the exact spectral shape also differs from the semi-analytic estimate in that it features an extended high frequency tail. As the black hole evaporates, it grows smaller and hotter, and the resultant gravitons emitted towards the end of the black hole lifetime have a higher initial frequency, leading to the formation of the high-frequency tail.

\subsection{Beyond Blackbody \& Instant Decay:\\Exact Numerical Results}

In order to obtain the exact instantaneous graviton spectrum, we use the publicly available code {\tt BlackHawk} \cite{BlackHawk:2019,BlackHawk:2021}, which goes beyond both the blackbody and instantaneous decay approximations. {\tt BlackHawk} uses tabulated and appropriately interpolated greybody factors to precisely compute the emission rates of Eq. (\ref{flux}) for all primary particle species. These are then used to solve for the black hole mass and angular momentum loss rates of Eqs. (\ref{massrate}) and (\ref{Jrate}) in order to obtain the overall black hole evolution. The result is time-dependent spectra which incorporate both greybody factors and the evolution of black hole mass and temperature. While {\tt BlackHawk} does allow for the study of black hole populations with extended mass and spin distributions, we presume monochromatic spectra for simplicity, as our primary focus will be on the effect of modified cosmological expansion histories.

We denote by $Q_{\rm GW}(t,\omega) \equiv \frac{dN_{\rm grav}}{dt d\omega}$ the instantaneous graviton flux, which is an output of {\tt BlackHawk}. The corresponding instantaneous power is $\frac{dE_{\rm grav}}{dt d\omega} = \frac{\omega}{2\pi} Q_{\rm GW}(t,\omega)$ and the instantaneous energy density emitted in the form of gravitational waves from an evaporating population of PBHs with number density $n_{\rm BH}(t)$ is:
\begin{equation}
    \frac{d\rho_{\rm GW}}{dt d\omega} = n_{\rm BH}(t) \frac{\omega}{2\pi} Q_{\rm GW}(t,\omega) \,.
\end{equation}
In order to obtain the total energy emitted in the form of gravitational waves, we need to integrate this quantity over the course of the black hole lifetime, from formation\footnote{Black holes formed during radiation domination tend to have negligible spin, and so the formation of near-extremal Kerr black holes will likely require the introduction of new physics. While it is possible to spin up a population of PBH through accretion and mergers, the maximal spin parameter obtained in this way is $\alpha_\star \sim 0.7$. Black holes that form from the collapse of density perturbations during a period of early matter domination, on the other hand, tend to have near-extremal spins \cite{Arbey:2021}.} at $t_i \simeq \frac{M}{M_{\rm Pl}^2}$ to evaporation at $t_*$, which are also outputs of {\tt BlackHawk}. The energy density per logarithmic frequency interval at evaporation then looks like
\begin{equation}
    \frac{d\rho_{\rm GW}^*}{d \ln \omega_*} = n_{{\rm BH},i} \left( \frac{a_i}{a_*} \right)^3 \frac{\omega_*^2}{2\pi} \int_{t_i}^{t_*} dt \, \frac{a_*}{a(t)} Q_{\rm GW}\big(t,\omega_* \frac{a_*}{a(t)} \big) \,,
\end{equation}
where, once again, we denote quantities evaluated at evaporation with a ``$*$". The exact time-dependence of the scale factor $a(t)$ appearing in this expression can be obtained by solving numerically the equations governing the background evolution
\begin{equation}\label{background}
\begin{split}
    & \frac{\dot{a}}{a} = \sqrt{\frac{8\pi}{3 M_{\rm Pl}^2} (\rho_{\rm BH} + \rho_{\rm rad})} \,,\\
    & \dot{\rho}_{\rm BH} + 3 \frac{\dot{a}}{a} \rho_{\rm BH} = \rho_{\rm BH} \frac{\dot{M}}{M} \,,\\
    & \dot{\rho}_{\rm rad} + 4 \frac{\dot{a}}{a} \rho_{\rm rad} = - \rho_{\rm BH} \frac{\dot{M}}{M} \,,
\end{split}
\end{equation}
where the overhead dot denotes a derivative with respect to coordinate time and the mass loss rate, given schematically in Eq. (\ref{massrate}), is a {\tt BlackHawk} output. To go from the spectrum at evaporation to that today, we redshift by four powers of the scale factor according to Eq. (\ref{redshifttotoday}), with $a_*$ again given by Eq. (\ref{astar}). Finally redshifting also the frequency to today as $\omega_0 = \omega_* a_*$ and dividing by the critical density, we arrive at the fractional energy density in gravitational waves today:
\begin{equation}\label{OmegaGW}
    \Omega_{\rm GW} = \frac{4 n_{{\rm BH},i} a_i^3}{3 M_{\rm Pl}^2 H_0^2} \omega_0^2 \int_{t_i}^{t_*} \frac{dt}{a(t)} Q_{\rm GW} \big( t,\frac{\omega_0}{a(t)} \big) \,.
\end{equation}

\begin{figure}[t]
\centering
\includegraphics[width=0.44\textwidth]{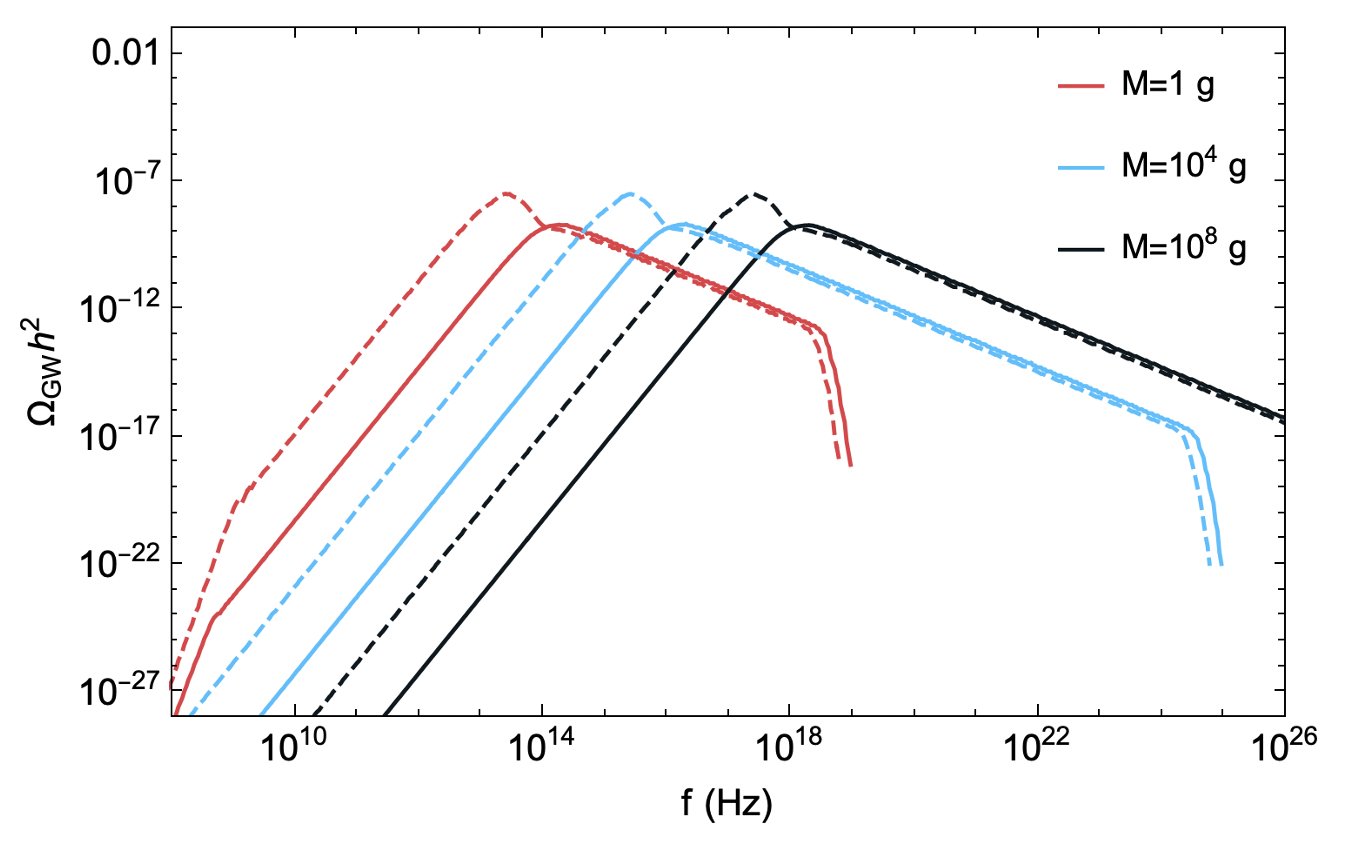}
\hspace{2mm}
\caption{Spectral density parameter $\Omega_{\rm GW} h^2$ for a monochromatic spectrum of PBHs of initial mass $M$ and mass fraction $\Omega_{{\rm BH},i}$, taken to be sufficiently large that the black holes come to dominate before decay. The peak amplitude is enhanced by several orders of magnitude for near-extremal rotating black holes ($\alpha_\star = 0.999$, dashed line) as compared with the non-rotating case ($\alpha_\star = 0$, solid line). Cosmological evolution is otherwise standard.}
\label{standardcosmo}
\end{figure}

In Fig. \ref{standardcosmo} we show the spectral shape of the energy density in gravitational waves today $\Omega_{\rm GW} h^2$ for a sampling of initial masses. In all cases we take the initial fractional energy in black holes $\Omega_{{\rm BH},i}$ sufficiently large that they come to dominate the energy density before decay, in which case the precise value of $\Omega_{{\rm BH},i}$ has no bearing on the spectrum. Solid lines correspond to the Schwarzschild case ($\alpha_\star=0$) while dotted lines correspond to Kerr black holes with near-extremal spin, $\alpha_\star = 0.999$. 

We see that the inclusion of spin has a significant impact on both the shape and amplitude of the gravitational wave spectrum today. First, note that for near-extremal black holes the peak position shifts to lower frequencies by about an order of magnitude. This is because rapidly spinning black holes have a reduced lifetime as compared with their non-spinning counterparts, and so the gravitational wave spectrum experiences a longer period of cosmological redshift\footnote{Based on the semi-analytic estimate, one might naively conclude that this shift towards lower frequencies is due to the fact that rapidly spinning black holes are considerably cooler than those of negligible spin, as can be seen from Eq. (\ref{temperature}). However for Kerr black holes, typical graviton energies are not of order the temperature, but rather the combination $T_{\rm BH}+m\Omega_{\rm BH}$, with $\Omega_{\rm BH}$ the black hole angular momentum given in Eq. (\ref{angmom}). The large value of $\Omega_{\rm BH}$ for rapidly spinning black holes compensates for the smaller $T_{\rm BH}$, such that the typical energies are comparable or even larger than those of gravitons from the analogous Schwarzschild black hole.}. The amplitude of the peak is also enhanced by several orders of magnitude due to the black hole spin, which acts as an effective chemical potential biasing the emission of higher spin particles. As was the case for our analytic estimate, we see that smaller mass black holes correspond to spectra that peak at lower frequencies. This is because they can be produced at earlier times and evaporate more quickly, leading to a longer period of cosmological redshift. 

We remark that the results of this section are in good agreement with \cite{Dong:2015}, which also considered the gravitational wave signal from evaporating Kerr black holes presuming standard cosmological evolution. They too found that the typical peak frequencies were much too high for detection in current and future gravitational wave experiments. In light of this, we now go beyond this analysis to consider the kinds of gravitational signals possible in non-standard cosmologies.

\section{Non-Standard Cosmologies}\label{sec:nonstandard}

\subsection{Early Matter Domination} 

Gravitational waves from evaporating PBHs can constitute an appreciable fraction of the energy density today. However, in most cases, the spectrum peaks at very high frequencies, far outside the reach of current and near-future detectors. Even for the best-case scenario of a population of Planck mass black holes which come to dominate the energy density before decay, the peak frequency is only as low as $\sim 10^{11}$ Hz. One way to effect a shift to lower frequencies would be to invoke a period of early matter domination (EMD). Since the universe expands to a greater extent in a fixed amount of time during matter domination than during radiation domination, this introduces extra cosmological redshift which serves to translate the spectrum to lower frequencies.

To induce this period of EMD, we introduce a heavy auxiliary field $\phi$ and demand that its initial energy density be greater than that in both black holes and radiation combined $\rho_{\phi,i}>\rho_{{\rm BH},i} + \rho_{{\rm rad},i}$. The universe will then remain matter dominated through black hole evaporation up until the time $\tau_\phi$ at which $\phi$ decays, replenishing the SM radiation bath and reheating the universe. To distinguish this from the situation of the previous section, this should occur after black hole evaporation has completed, $\tau_\phi > t_*$. Note that while it is true that the black holes themselves will generically serve to induce a transient period of EMD (presuming they possess sufficient initial energy density and lifetime), this ends at the time of evaporation. If instead EMD comes about as a result of a heavy auxiliary field, then it can last much longer --- potentially up until $\sim \mathcal{O}(1)$ s, when radiation domination must begin so as not to spoil the predictions of BBN.

An added benefit to using the auxiliary field to establish EMD is that the black holes can then form during the matter dominated epoch with appreciable spin. For the standard cosmology case of the section prior, we were agnostic as to how the black hole population acquired near-extremal spins. Given that black holes formed from the collapse of overdensities during radiation domination generally have negligible spin, one would have to rely on accretion and mergers, which generally only result in a spin parameter of $\alpha_\star \lesssim 0.7$. On the other hand, black holes formed during matter domination typically have appreciable spin, which is ideal in terms of maximizing the amplitude of the GW signal.

In the typical radiation dominated case, a density perturbation will collapse to form a black hole upon re-entering the horizon if its amplitude is greater than some threshold value $\delta_{\rm th}$. Naively applying the analytic formula for the threshold in the matter dominated case, one finds that it vanishes $\delta_{\rm th} \rightarrow 0$, which would seem to suggest that any region of overdensity could collapse to a black hole. This is of course not the case, as it turns out that non-spherical effects play a crucial role for collapse during matter domination. As an overdensity begins to collapse, its angular momentum grows significantly and prevents collapse in a majority of cases. Those horizons that do succeed in collapsing, however, form black holes which are rapidly spinning with near-extremal values of $\alpha_\star \sim 1$. See e.g. \cite{Harada:2016,Harada:2017,Arbey:2021} for further detail regarding the computation of the initial mass fraction $\Omega_{{\rm BH},i}$ in the matter dominated case.

In terms of the gravitational wave signal $\Omega_{\rm GW}$, the derivation proceeds as in the case of standard cosmological evolution, and the spectral density parameter today is still given by Eq. (\ref{OmegaGW}) with $n_{{\rm BH},i}$ in Eq. (\ref{nBHi}). The difference\footnote{Technically the introduction of a new species changes the particle emission rate from the black hole, but the effect is completely negligible. We could also take $\phi$ to be sufficiently heavy such that its emission from the black hole is statistically suppressed.} lies in the input for the initial conditions as well as the evolution of the scale factor $a(t)$ itself. With the addition of the auxiliary field $\phi$, Eqs. (\ref{background}) governing the background evolution become
\begin{equation}
\begin{split}
    & \frac{\dot{a}}{a} = \sqrt{\frac{8\pi}{3 M_{\rm Pl}^2}(\rho_{\phi} + \rho_{\rm BH} + \rho_{\rm rad})} \,,\\
    & \dot{\rho}_{\phi} + 3 H \rho_\phi = - \Gamma_\phi \rho_\phi \,, \\
    & \dot{\rho}_{\rm BH} + 3 H \rho_{\rm BH} = \frac{\dot{M}}{M} \rho_{\rm BH} \,, \\
    & \dot{\rho}_{\rm rad} + 4 H \rho_{\rm rad} = \Gamma_\phi \rho_\phi - \frac{\dot{M}}{M} \rho_{\rm BH} \,,
\end{split}
\end{equation}
where the decay rate is approximately $\Gamma_\phi \simeq \frac{m_\phi^3}{M_{\rm Pl}^2}$ on dimensional grounds. The mass of $\phi$ should be appropriately chosen to ensure the corresponding lifetime $\tau_\phi = \Gamma_\phi^{-1}$ is after evaporation but before BBN. 

\begin{figure}[t]
\centering
\includegraphics[width=0.44\textwidth]{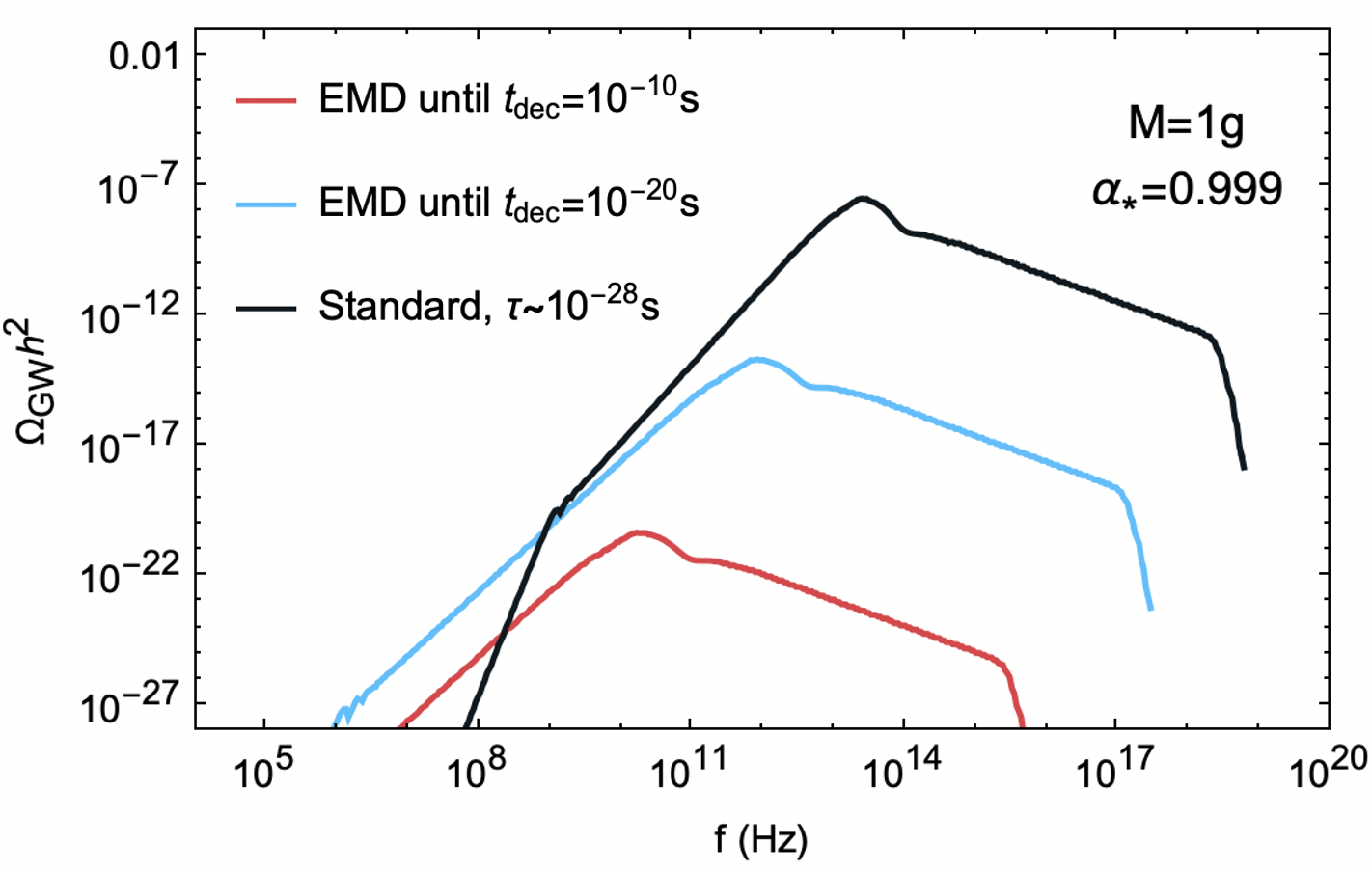}
\hspace{2mm}
\includegraphics[width=0.44\textwidth]{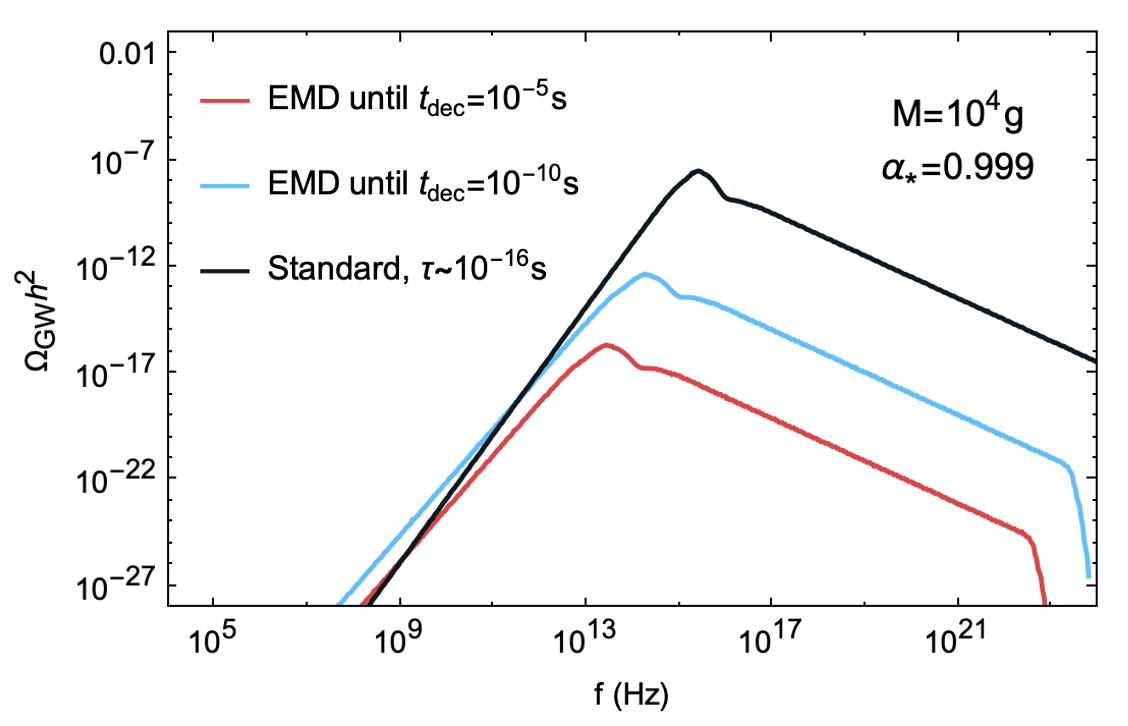}
\hspace{2mm}
\caption{Spectral density parameter $\Omega_{\rm GW} h^2$ presuming black hole formation and evaporation during a period of early matter domination induced by the presence of a heavy auxiliary field $\phi$ which lasts until a time $t_{\rm dec}$.}
\label{EMDplots}
\end{figure}

Solving this system of equations and evaluating Eq. (\ref{OmegaGW}) for the spectrum today for various choices of decay time $t_{\rm dec} = t_i+\tau_\phi$ yields the plots in Fig. \ref{EMDplots}. We see that a longer period of EMD is associated with a more heavily redshifted spectrum, as expected. For the best case scenario of a population of Planck mass black holes, letting EMD persist until $\sim 1$ s can bring the peak frequency as low as $\sim 10^4$ Hz. However, the associated signal strength is vanishingly small at $\Omega_{\rm GW} h^2 \sim 10^{-36}$. Unfortunately, this is a generic side effect of the energy density in gravitational waves falling off as $\rho_{\rm GW} \sim a^{-4}$ while the frequency decreases as $f \sim a^{-1}$. 

\subsection{Generalized Equation of State}

Extra cosmological redshift coming from a period of early matter domination, or any period with equation of state $w<1/3$, will result in lower peak frequencies but also dramatically smaller amplitudes, as previously noted. Given that the peak signal is barely on the cusp of observability $\Omega_{\rm GW} \sim 10^{-7.5}$ in the standard scenario, any period of slower expansion which serves to dilute energy density will lower the signal to outside of sensitivity. 

In this section, we consider how the signal from graviton production changes upon varying the equation of state $w$ of a species that dominates the energy density of the universe at early times.

We first note that the gravitational wave signal is not the only potential observable associated with graviton emission from PBHs. High energy gravitons emitted by a population of light, evaporating PBH constitute a thermal background of dark radiation, which contributes to the effective number of neutrino species $N_{\rm eff}$, defined as the contribution to the radiation energy density beyond that of photons:
\begin{equation}
    \rho_{\rm rad} = \rho_\gamma \left( 1 + \frac{7}{8} \left( \frac{4}{11} \right)^{4/3} N_{\rm eff} \right) \,.
\end{equation}
It is useful to factor this as $N_{\rm eff} = N_{\rm eff}^{\rm SM} + \Delta N_{\rm eff}$, where $N_{\rm eff}^{\rm SM} = 3.046$ accounts for the contribution from SM neutrinos and $\Delta N_{\rm eff}$ parameterizes the departure from the SM prediction. This is tightly constrained by both BBN and CMB measurements. In particular, the one-tailed \textit{Planck} TT, TE, EE+lowE+lensing+BAO constraint is $\Delta N_{\rm eff} < 0.30$ at 95\% \cite{Planck:2018}. The contribution to $\Delta N_{\rm eff}$ from gravitational waves reads:
\begin{equation}
    \Delta N_{\rm eff} = \frac{8}{7} \left( \frac{11}{4} \right)^{4/3} \frac{\rho_{\rm GW}}{\rho_\gamma} \,.
\end{equation}

The $\Delta N_{\rm eff}$ bound is an integral bound which applies to the total energy density in gravitational waves integrated over all frequencies. It can thus be interpreted as a bound on the maximum amplitude of the GW spectrum, $\Omega_{\rm GW}^{\rm max}$. Then from $\rho_{\rm GW} \simeq \Omega_{\rm GW}^{\rm max} \rho_{\rm crit}$ and $\rho_\gamma = \frac{\pi^2}{15} T_0^4$, with $T_0 = 0.235\, \text{meV}$, the CMB temperature today, we find:
\begin{equation}
    \Delta N_{\rm eff} \simeq \frac{120}{7 \pi^2} \left( \frac{11}{4} \right)^{4/3} \frac{\rho_{\rm crit}}{T_0^4} \Omega_{\rm GW}^{\rm max} \,.
\end{equation}
Demanding that $\Delta N_{\rm eff} \lesssim 0.30$ translates to a constraint on the spectral density parameter:
\begin{equation}\label{Neffbound}
    \Omega_{\rm GW}^{\rm max} \lesssim 3.6 \times 10^{-6} \,.
\end{equation}

The possibility that the early universe might be dominated by a species whose energy density redshifted differently from radiation would modify the thermal history of the universe and affect both the spectrum of gravitons emitted by evaporating PBHs and the effective number of relativistic degrees of freedom. We shall consider here cosmological models where at early times the energy density is dominated by a species $\phi$ with a {\em generalized} equation of state $P_\phi=w_\phi \rho_\phi$ and $w_\phi>1/3$, corresponding to a period of faster expansion relative to the standard case. If black hole evaporation occurs before or during $\phi$ domination, then the energy density in gravitational waves will experience less dilution, potentially giving rise to a signal which saturates the $\Delta N_{\rm eff}$ bound, and which might even be detectable with future, high-frequency gravitational wave searches, as discussed below.

The energy density in $\phi$ evolves with the scale factor $a$ as
\begin{equation}\label{phiscaling}
    \rho_\phi\sim a^{-(4+n)} \,,
\end{equation}
where we follow the same notation as \cite{DEramo:2017a,DEramo2017b} and parameterize the deviation from the radiation scaling with $n=3 w_\phi -1$. Generically, for a canonically normalized scalar field minimally coupled to gravity, the equation of state is $w_\phi =(K-V)/(K+V)$, where
\begin{equation}
    K=\frac{1}{2}\left(\frac{{\rm d}\phi}{{\rm d}t}\right)^2,
\end{equation}
and $V=V(\phi)$ is the potential. Then $w_\phi \to 1$ for $K\gg V$ (a regime dubbed {\em kination}), and $w_\phi \to -1$ for $K \ll V$, yielding a range of $-4 \le n \le 2$. Models with $n>2$ are also possible\footnote{Notice that for $n>2$ there is no causality violation, despite having $p_\phi>\rho_\phi$, since the sound speed is $c^2_s=1$ \cite{Armendariz-Picon:1999,Christopherson:2008}.} for instance in the context of  ekpyrotic scenario \cite{Khoury:2001} or with periodic potentials and a varying $w_\phi$ \cite{Choi:1999,Gardner:2004}.

\begin{figure}[t]
\centering
\includegraphics[width=0.43\textwidth]{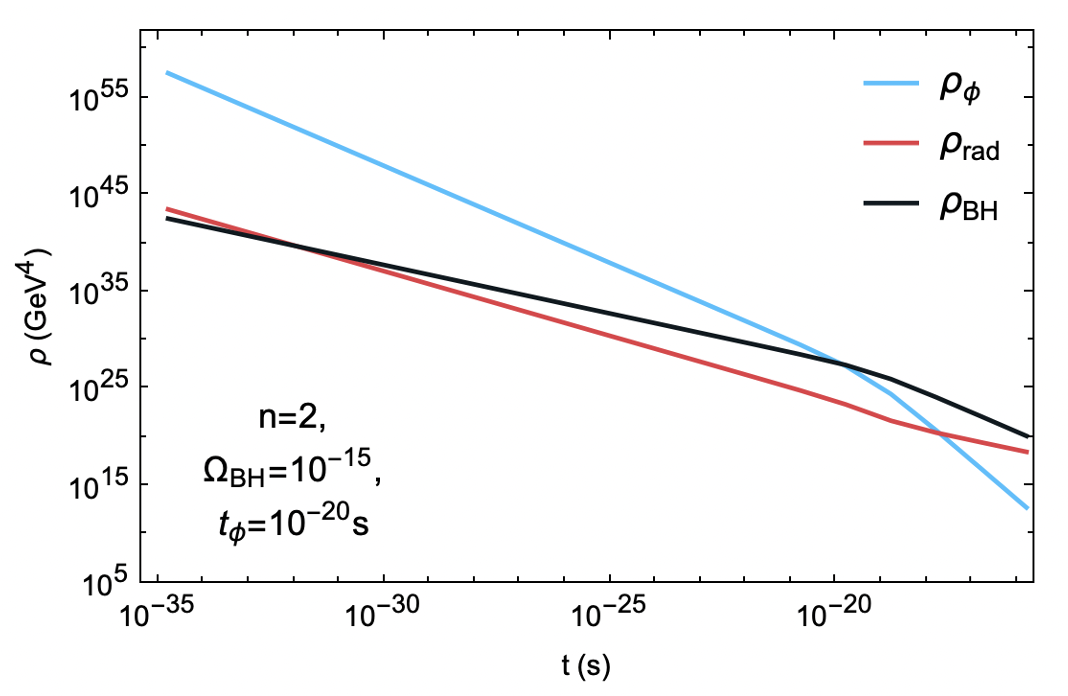}
\hspace{2mm}
\caption{Sample evolution of $\rho_{\rm BH}$, $\rho_{\rm rad}$, and $\rho_\phi$ as obtained by solving  Eq. (\ref{nonstandardboltzmann}) for a population of near extremal ($\alpha_\star = 0.999$) PBH with initial mass $M_i=10^4$ g. We take $n=2$, such that the initial $\phi$ domination corresponds to kination. This continues until $\rho_{\phi}$ becomes subdominant to $\rho_{\rm BH}$ at a time $t_\phi = 10^{-20} \, \text{s}$. Black hole domination then continues until evaporation replenishes the radiation bath at $t_* \simeq 3 \times 10^{-15}$ s.}
\label{backgroundevo}
\end{figure}

The background evolution is described by the following equations:
\begin{equation}\label{nonstandardboltzmann}
\begin{split}
    & \frac{\dot{a}}{a} = \sqrt{\frac{8\pi}{3 M_{\rm Pl}^2}(\rho_{\phi} + \rho_{\rm BH} + \rho_{\rm rad})} \,,\\
    & \dot{\rho}_{\phi} + (4 + n) H \rho_\phi = 0 \,, \\
    & \dot{\rho}_{\rm BH} + 3 H \rho_{\rm BH} = \frac{\dot{M}}{M} \rho_{\rm BH} \,, \\
    & \dot{\rho}_{\rm rad} + 4 H \rho_{\rm rad} = - \frac{\dot{M}}{M} \rho_{\rm BH} \,,
\end{split}
\end{equation} 
with $n>0$. We specify the initial fractional energy densities $\Omega_{{\rm BH},i}$, $\Omega_{\phi,i}$, and $\Omega_{{\rm rad},i}=1-\Omega_{{\rm BH},i} - \Omega_{\phi,i}$ at the time of black hole formation, where $\Omega_{x,i} = \rho_{x,i}/\rho_i$ and:
\begin{equation}
    \rho_i = \frac{3 M_{\rm Pl}^2}{2(4+n)^2 \pi} \frac{1}{t_i^2} \,.
\end{equation}
Note that $\phi$ will generically dominate at early times by virtue of the way its energy density scales with redshift, Eq. (\ref{phiscaling}), giving rise to an expansion with $a \sim t^{2/(4+n)}$. We take the PBH to form during this period at a time $t_i = \frac{4}{4+n} \frac{M_i}{M_{\rm Pl}^2}$. As the universe expands, the energy density in $\phi$ quickly becomes subdominant, as demonstrated in Fig. \ref{backgroundevo}, which shows the evolution of the energy densities of the various components for different choices of initial conditions. We denote by $t_\phi$ the time at which the energy density in $\phi$ becomes subdominant to that in radiation and black holes, $\rho_\phi(t_\phi) = \rho_{\rm rad}(t_\phi) + \rho_{\rm BH}(t_\phi)$.

One might wonder about the efficiency of PBH formation during a period of equation of state greater than that of radiation, $w>1/3$. The authors of \cite{Harada:2013} have argued that the threshold for overdensity collapse in general cosmologies with $w \geq 0$ ($n \geq -1$) is given by the following analytic formula\footnote{This nominally vanishes for the matter dominated case $n=-1$, which would seem to suggest any overdensity should collapse to a black hole. However the derivation of \cite{Harada:2013} posits spherical symmetry, and as we have already argued, deviations from spherical symmetry and angular momentum play a large role in the suppression of PBH formation during matter domination.}
\begin{equation}
    \delta_{\rm th} = \left( \frac{n+4}{n+6} \right) \sin^2 \left( \frac{\pi}{\sqrt{3}} \frac{\sqrt{n+1}}{(n+2)} \right) \,,
\end{equation}
where the prefactor assumes comoving gauge. This is based on the Jeans criterion as the determining factor for PBH formation and is derived by demanding that the free-fall timescale be shorter than the soundwave propagation timescale, such that gravitational collapse wins out over the pressure gradient. For radiation domination $n=0$ it gives $\delta_{\rm th} \simeq 0.414$ while for kination $n=2$ we have $\delta_{\rm th} \simeq 0.375$. Thus forming PBH during an early period with $n>0$ should be comparable if not marginally easier when compared with the radiation dominated case.

\begin{figure}[t]
\centering
\includegraphics[width=0.43\textwidth]{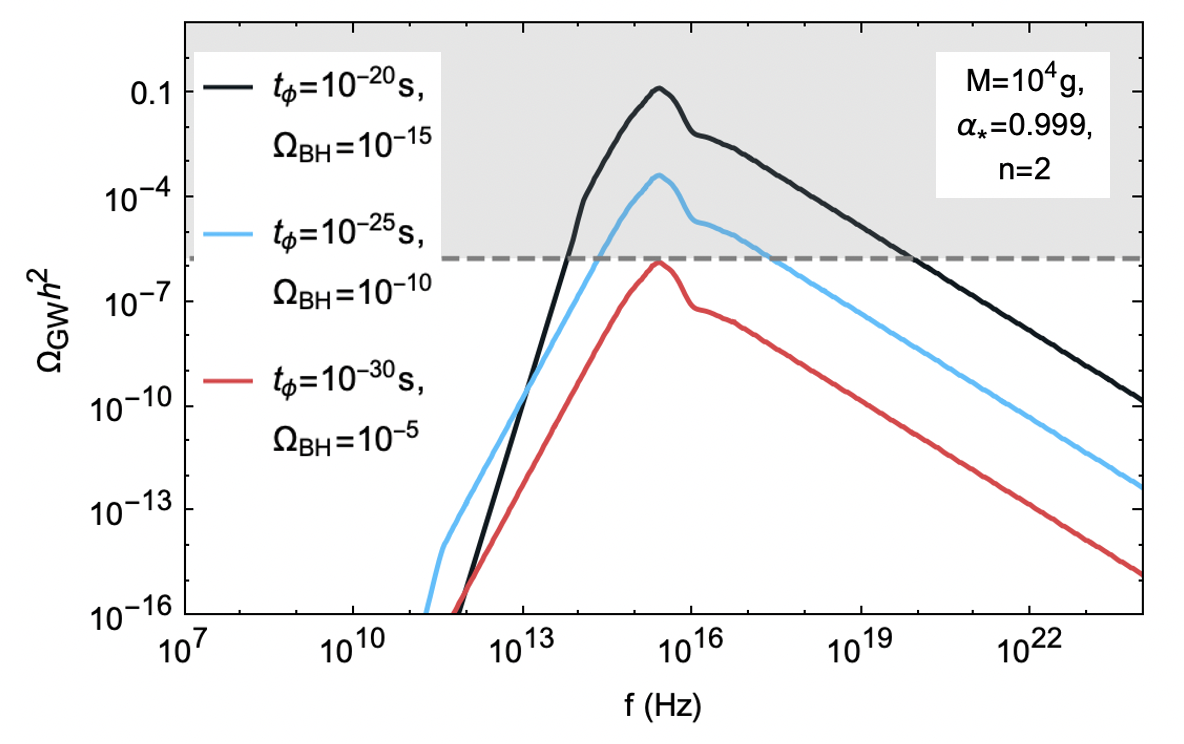}
\hspace{2mm}
\caption{Spectral density parameter $\Omega_{\rm GW} h^2$ as compared with the $\Delta N_{\rm eff}$ bound of Eq. (\ref{Neffbound}) (grey shaded region) presuming $\phi$ domination until $t_\phi$. Longer periods of kination stemming from larger initial energy density in $\phi$ result in an amplified gravitational wave signal, potentially contributing inappropriately to $\Delta N_{\rm eff}$.}
\label{Neffplot}
\end{figure}

Ignoring for now more exotic scenarios with $n>2$, the gravitational wave signal will be greatest for the case of kination $n=2$, during with the energy density redshifts as $\rho \sim a^{-6}$. In such a scenario it is easy to violate the bound on $\Delta N_{\rm eff}$, as shown in Fig. \ref{Neffplot}.
The duration of kination is fixed by the choice of initial conditions for the fractional energy density in each sector, with a larger initial density corresponding to a longer period of kination (see Fig. \ref{backgroundevo}). For our benchmark points, we choose $\Omega_{\phi}$ which gives the desired $t_\phi$ and set $10\%$ of the remaining energy density to be in the form of PBH, $\Omega_{\rm BH} = 0.1 (1 - \Omega_{\phi})$. We see that even an extremely transient period of kination, completing long before black hole evaporation completes, can result in a largely boosted signal, which can be ruled out on the grounds of violation of $\Delta N_{\rm eff}$ bounds.

Fig.~\ref{Neffplot} studies, for a mass $M=10^4$ grams, and quasi-maximal Kerr ($a_*=0.999$), the spectrum of gravitational waves emitted in kination domination, with differing times $t_\phi$ at which the kination energy density equals the energy density of the other components, $t_\phi=10^{-30}$ s (red line), $10^{-25}$ s (blue line), and $10^{-20}$ s (black line). The density of gravitational waves increases the longer the period of kination domination, with the peak well inside the region excluded by $\Delta N_{\rm eff}$. We also observe a steepening of the lower-frequency tail (below the peak), while the high-frequency behavior does not depend on the $t_\phi$.

\section{Discussion, Observational Prospects, and Conclusions}\label{sec:conclusions}
While the general upper bound on the gravitational wave spectrum from evaporating PBHs discussed above prevents the possibility to detect a stochastic gravitational wave signal in a standard cosmological scenario, this conclusion is affected by considering generalized early universe cosmologies, as detailed in the previous section. In particular, for $w_\phi \geq 1$ the gravitational wave amplitude can significantly exceed the $\Delta N_{\rm eff}$ bound, and potentially be detectable with future high-frequency gravitational wave searches.

The detection of ultra-high-frequency ($f\gg 1\ {\rm kHz}$) gravitational waves is an active area of intense experimental investigation (see e.g. the recent review \cite{Aggarwal:2020}). Several experimental techniques have been proposed, ranging from table-top interferometers, holometers, optically levitated sensors, devices based on the inverse Gertsenshtein effect (the conversion of gravitational waves to photons \cite{Gertsenshtein:1962kfm}), on gravitational wave to electromagnetic wave conversion in an electric or magnetic field, bulk acoustic-wave devices, superconducting rings, and graviton-magnon resonance. 

We consider the following broad classes of experimental techniques, in order of increasing frequency sensitivity, and with a proposed dimensionless sensitivity figure.\footnote{Note that we present our results in terms of the dimensionless characteristic strain $h_c$, related to the spectral density parameter as: 
\begin{equation*}
    \Omega_{\rm GW} = \frac{4 \pi^2}{3 H_0^2} f^2 h_c^2 \, .
\end{equation*}
}
\begin{itemize}
    \item Laser interferometers (1-10) kHz [9$\times10^{-26}$] \cite{Ackley:2020atn}
    \item Optically levitated sensors (10-100) kHz [4$\times10^{-24}$]  \cite{Aggarwal:2020}
    \item Enhanced magnetic conversion ($\sim$10) GHz [$10^{-30}$]  \cite{Li:2009zzy,Ito:2019wcb}
    \item Inverse Gersenshtein effect (10$^{14}$-10$^{18}$) Hz [3$\times10^{-30}$] \cite{Ejlli:2019bqj}
\end{itemize}
In particular, a promising technology repurposes axion-like particle conversion in a magnetic field to look for graviton conversion \cite{Ejlli:2019bqj}; here we highlight current constraints, and the associated relevant frequency range, for JURA  (Joint Undertaking on
the Research for Axion-like particles) \cite{Beacham:2019nyx}, OSQAR II \cite{OSQAR:2015qdv}, and CAST \cite{CAST:2017uph}.

\begin{figure*}[t]
\centering
\includegraphics[width=\textwidth]{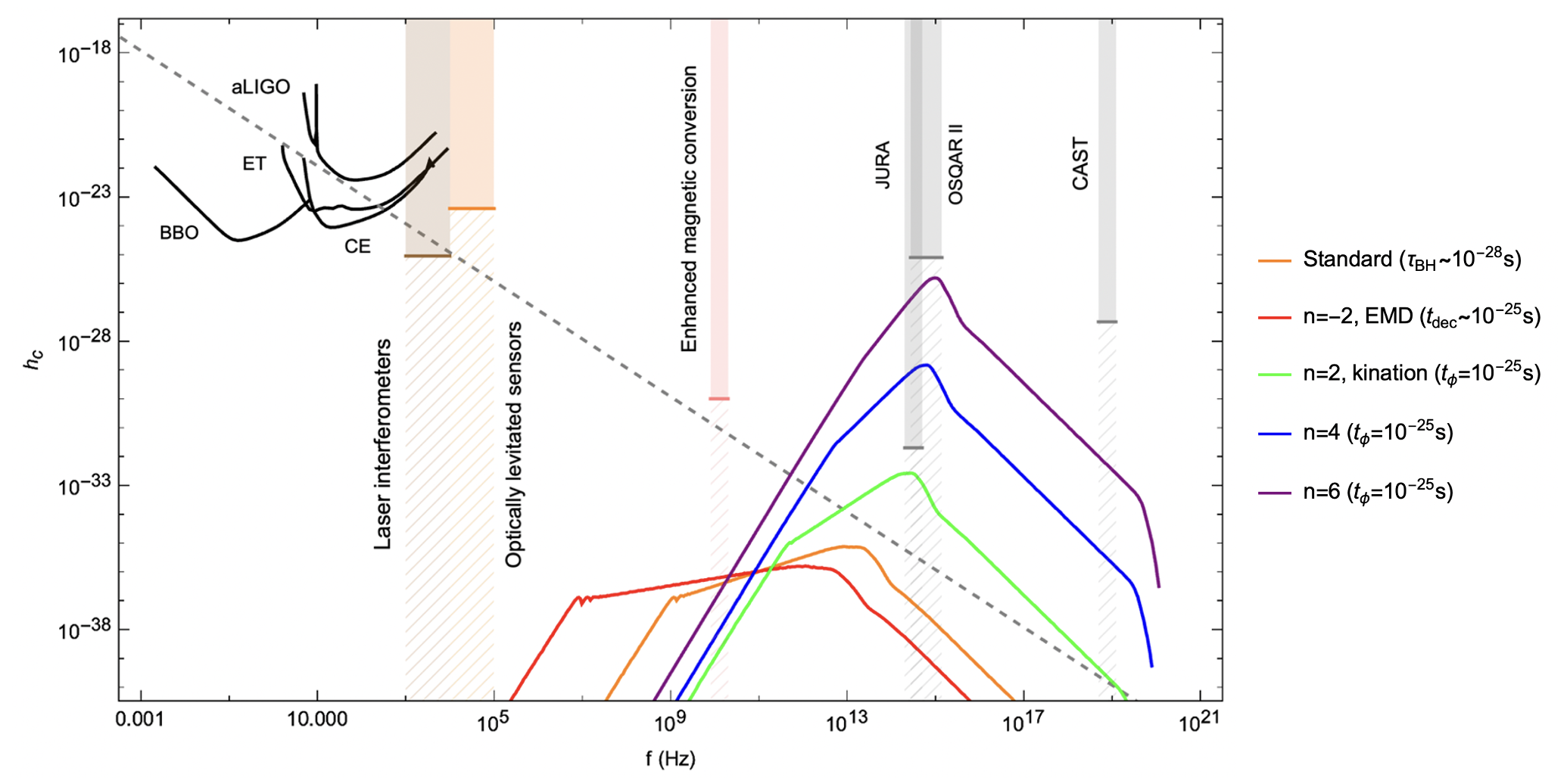}
\hspace{2mm}
\caption{Characteristic strain $h_c$ of $M=1$ g, quasi-extremal ($\alpha_\star = 0.999$) PBHs for a sample of early universe cosmologies  compared with the sensitivity of several proposed high-frequency gravitational wave detector technologies, as well as the $\Delta N_{\rm eff}$ bound from \textit{Planck} \cite{Planck:2018} (grey, dashed).}
\label{sensitivityplot}
\end{figure*}
Fig.~\ref{sensitivityplot} compares existing proposals for high-frequency gravity wave detectors with our predictions for the gravitational wave emission from PBH evaporation with standard and non-standard early-universe cosmological histories. The latter are taken for the benchmark case of $M=1$ g, $\alpha_\star = 0.999$ PBH and shown as colored lines: The standard prediction, featuring PBH evaporation at $\tau_{\rm BH} \sim 10^{-28}$ s, is shown in orange. The early matter domination case is shown in red, and assumes that the ``extra'', non-standard species (in this case, a non-relativistic matter component) decays at $t_{\rm dec} = 10^{-25}$ s. The green line corresponds to a kination scenario where the species responsible for kination becomes subdominant at $t_\phi=10^{-25}$ s; the blue and purple lines corresponds to an even-faster redshifting species with $n=4,\ 6$, also becoming subdominant at the same time, $t_\phi=10^{-25}$ s.

Our general findings are that for large $n$ (i.e. for a ``stiff'' equation of state, $w_\phi\ge 1$), the peak gravitational wave emission can be quite ``bright'', exceeding limits from the number of relativistic species (dashed grey line), and well into the frequency range of future high-frequency gravitational wave detectors; we find that the very high and very low frequency behavior of the gravitational wave spectrum is unchanged, but at frequency around and below the peak frequency, the different redshifting in different early universe cosmologies produces different spectral shapes. Finally, albeit for larger $w$, the spectrum shifts to higher frequency, we find that the peak spectrum for $n=2, 4, 6$ only mildly moves to higher frequencies.

In summary, we have considered the spectrum of gravitational waves produced by the evaporation of light, primordial black holes in the early universe in the context of generic cosmological histories. We first discussed the general features of the signal in a standard cosmological setting where the PBH form and evaporate in radiation domination, potentially with a brief period of black hole domination, and highlighted how there is a general upper limit to the intensity of the ensuing gravitational wave stochastic background. We then studied the case of early matter domination by a species different from the PBH themselves, and concluded that the peak of the gravity wave spectrum shifts to lower frequencies, but is also significantly suppressed in intensity, leading to bleak detection prospects. Finally, we entertained scenarios where the early universe is dominated by a species redshifting {\em faster} than radiation, such as kination of super-stiff fluids. In those cases, while the peak gravitational wave emission is shifted to {\em higher} frequencies, the intensity of the peak emission is also greatly enhanced, and possibly in conflict with constraints from the number of relativistic degrees of freedom. On a more optimistic note, however, this offers opportunities for discovery for future high-frequency gravitational wave detectors.\\

\section*{Acknowledgements}
S.P. and J.S. are partly supported by the U.S. Department of Energy grant number de-sc0010107. 

\bibliography{references}

\end{document}